\documentclass[conference]{IEEEtran}
\IEEEoverridecommandlockouts
\usepackage{cite}
\usepackage{amsmath,amssymb,amsfonts}
\usepackage{algorithm}
\usepackage{algorithmic}
\usepackage{graphicx}
\usepackage{textcomp}
\usepackage{xcolor}
\usepackage{makecell}
\usepackage{multirow}

\usepackage[OT1]{fontenc}
\usepackage{caption}
\usepackage{booktabs} %需要加载宏包{booktabs}
\def\BibTeX{{\rm B\kern-.05em{\sc i\kern-.025em b}\kern-.08em
    T\kern-.1667em\lower.7ex\hbox{E}\kern-.125emX}}

\begin{document}
\title{FGAM:Fast Adversarial Malware Generation Method Based on Gradient Sign\\}

\author{\IEEEauthorblockN{1\textsuperscript{st}  Kun Li}
    \IEEEauthorblockA{
        \textit{PLA Information Engineering University}\\
        Zhengzhou, China \\
        moyue\_lk@foxmail}
    \and
    \IEEEauthorblockN{2\textsuperscript{nd}  Fan Zhang}
    \IEEEauthorblockA{
        \textit{PLA Information Engineering University}\\
        Zhengzhou, China \\
    }
    \and
    \IEEEauthorblockN{3\textsuperscript{rd}  Wei Guo}
    \IEEEauthorblockA{
        \textit{PLA Information Engineering University}\\
        Zhengzhou, China \\
    }
}

\maketitle

\begin{abstract}
    Malware detection models based on deep learning have been widely used, but recent research shows that deep learning models are vulnerable to adversarial attacks. Adversarial attacks are to deceive the deep learning model by generating adversarial samples. When adversarial attacks are performed on the malware detection model, the attacker will generate adversarial malware with the same malicious functions as the malware, and make the detection model classify it as benign software. Studying adversarial malware generation can help model designers improve the robustness of malware detection models. At present, in the work on adversarial malware generation for byte-to-image malware detection models, there are mainly problems such as large amount of injection perturbation and low generation efficiency. Therefore, this paper proposes FGAM (Fast Generate Adversarial Malware), a method for fast generating adversarial malware, which iterates perturbed bytes according to the gradient sign to enhance adversarial capability of the perturbed bytes until the adversarial malware is successfully generated.
    It is experimentally verified that the success rate of the adversarial malware deception model generated by FGAM is increased by about 84\% compared with existing methods.

\end{abstract}

\begin{IEEEkeywords}
    Adversarial malware, Adversarial examples, Adversarial attacks, Deep learning, Malware detection
\end{IEEEkeywords}

\section{Introduction}
Malware is the primary method of network attacks, which can infect users' computers without the user's permission, posing considerable threats to users' information and property security\cite{agrawal2014evaluation}. With the rapid development of Internet technology, the cost of spreading malware is further reduced. In recent years, the amount of malware has proliferated. To deal with the rapidly growing malware, researchers began to use deep learning techniques to detect malware\cite{raff2017malware}\cite{liu2021mg}\cite{li2022intelligent}. Traditional malware detection methods\cite{ye2017survey} detect malware by manually setting filter rules. There are problems such as detection rules being too complex, detection rule setting requires a lot of expert knowledge, and detection rules need to be updated in real time. Different from it, the detection method based on deep learning automatically obtains malware features through data learning. Research shows that this detection method can have a high accuracy rate of malware detection, and does not rely on the quality of manual rules.

However, deep learning is vulnerable to adversarial examples due to the uneven distribution of training data and insufficient robustness of model design\cite{szegedy2013intriguing}. Adversarial attack was first proposed by Szegedy et al. \cite{szegedy2013intriguing}, and it is defined as fooling the neural network model by generating pair samples. In the work of Szegedy et al. \cite{szegedy2013intriguing}, perturbations are added to the pixels of the original image to generate adversarial images (adversarial examples). In addition, the perturbation size is limited to ensure that the original image is consistent with the human visual observations of the adversarial image. When the adversarial image (adversarial example) and the original image are input to the neural network, the neural network outputs different results. Adversarial attacks in malware are different from images. Malware has semantics, so perturbations must be limited to ensure that the generated adversarial malware has the same functionality as the original malware. In addition, adversarial malware should be able to fool the detection model, i.e., adversarial malware is classified as benign software by the detection model.

The malware detection model based on byte-to-image was first proposed by Cui et al. \cite{cui2018detection}, and it was further improved by Tekerek et al. \cite{tekerek2022novel}, and now it can have a high detection accuracy rate. The method first converts malware bytes into images, which are then classified using convolutional neural networks. Note that the malware detection model based on byte-to-image is also vulnerable to adversarial malware. However, there are some limitations to the adversarial malware generative approach to this model due to the discrete properties and semantics of PE(Portable Executable) files. Such as the generated adversarial malware losing their original functions\cite{liu2019atmpa}, the generation efficiency of the adversarial malware being low\cite{mao2022adversarial}\cite{benkraouda2021attacks}, and the injection of the generated adversarial malware being too much perturbation\cite{khormali2019copycat}. Therefore, in this paper, FGAM (Fast Generate Adversarial Malware) is proposed, a method for rapidly generating adversarial malware based on gradient signs, which injects perturbation through functionality-preserving manipulations to ensure that the adversarial malware has the same function as the original malware. In addition, the method iterates byte perturbation to enhance perturbation adversarial capability. The enhanced adversarial capability of byte perturbation can reduce the amount of perturbation and improve the success rate of generating adversarial malware. The main contributions of this paper are as follows:
\begin{enumerate}
    \item We use a reverse gradient sign to enhance its adversarial capabilities with iteration byte perturbation, and use function-preserving manipulations to inject perturbations to generate adversarial malware. In addition, we use the least square method to detect whether there is an oscillation in the malicious fraction drop rate of the model output, and shorten the process of generating adversarial malware by ending the oscillation process early.
    \item We calculate the entropy distribution of the original sample and adversarial malware and find that the perturbation injection ratio is limited, and the perturbation injection ratio and the hiding ability of adversarial malware are mutually restrictive.
    \item We conduct the transfer experiment of adversarial malware, and analyze the transfer attack capability of adversarial malware under different perturbation injection ratios, perturbation injection locations, and perturbation adversarial strengths.
\end{enumerate}

In this paper, we focus on the malware detection model based on byte-to-image and propose a method to generate adversarial malware quickly. In Section 2, we introduce the research background of adversarial malware and related work. In Section 3, the adversarial malware generation method proposed in this paper is presented. In Section 4, the effectiveness of the method in this paper is verified through experiments. Finally, the work of this paper is concluded, and future research directions are proposed in Section 5.

\section{Background and Related Work}
\subsection{Background}
\subsubsection{PE format}
\begin{figure}[hbtp]
    \centering
    \includegraphics[scale=0.5]{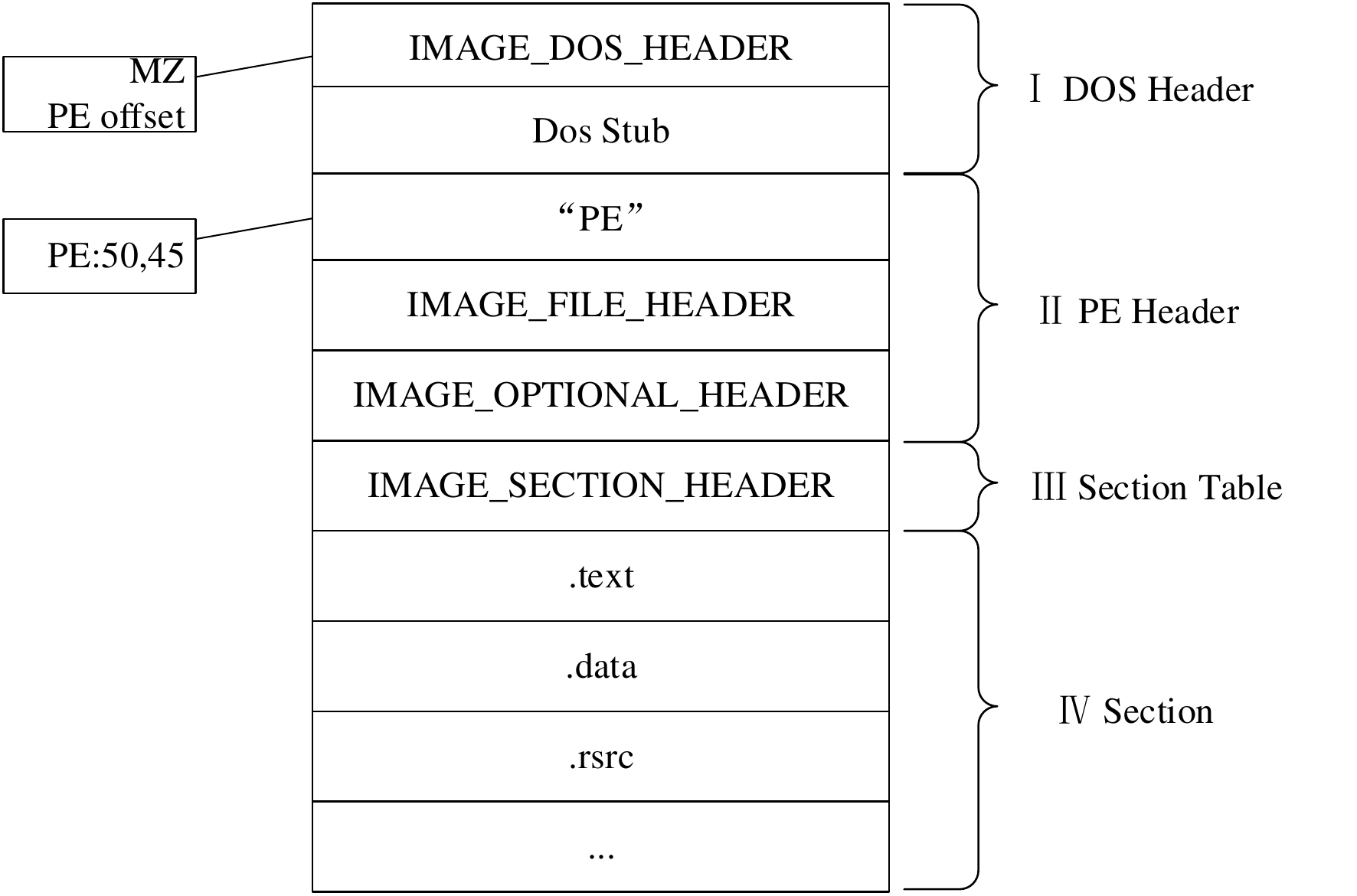}
    \caption{The Windows PE file format}
    \label{fig1}
\end{figure}

The structure of PE files in the Windows system is shown in Figure\ref{fig1}\footnote{https://docs.microsoft.com/en-us/windows/win32/debug/pe-format}, which is mainly composed of DOS header, PE header, section table and section. (1) DOS header: This contains the mark MZ of the PE file and the offset of the PE header. In addition, there is the DOS stub part, which is the data required for the PE file to be loaded in the DOS environment. (2) PE header: It contains the mark character "PE" of the PE header, and its corresponding hexadecimal is 50, 45. Also contains file headers, optional headers, etc. (3) Section table: section list, including the size, location, attributes and other information of the section (4) Section: byte block, the main content of the PE file, which stores the text, data, image and other information of the PE file.

The structure of the PE file has redundancy and alignment features. The redundancy feature is to ensure the compatibility of the PE file with the old system, and the alignment feature is to provide the high efficiency of the software when running. Due to the structural characteristics of PE files, this paper can inject content into PE files through functionality-preserving manipulations\cite{zaidan2009new}, and maintain the executable and functional properties of PE files. In addition, many deep learning-based malware detection methods utilize the PE file structure, which can easily extract static feature data such as API calls and import function tables from PE files.

\subsubsection{Deep learning-based malware detection}
\begin{figure*}[hbtp]
    \centering
    \includegraphics[scale=0.9]{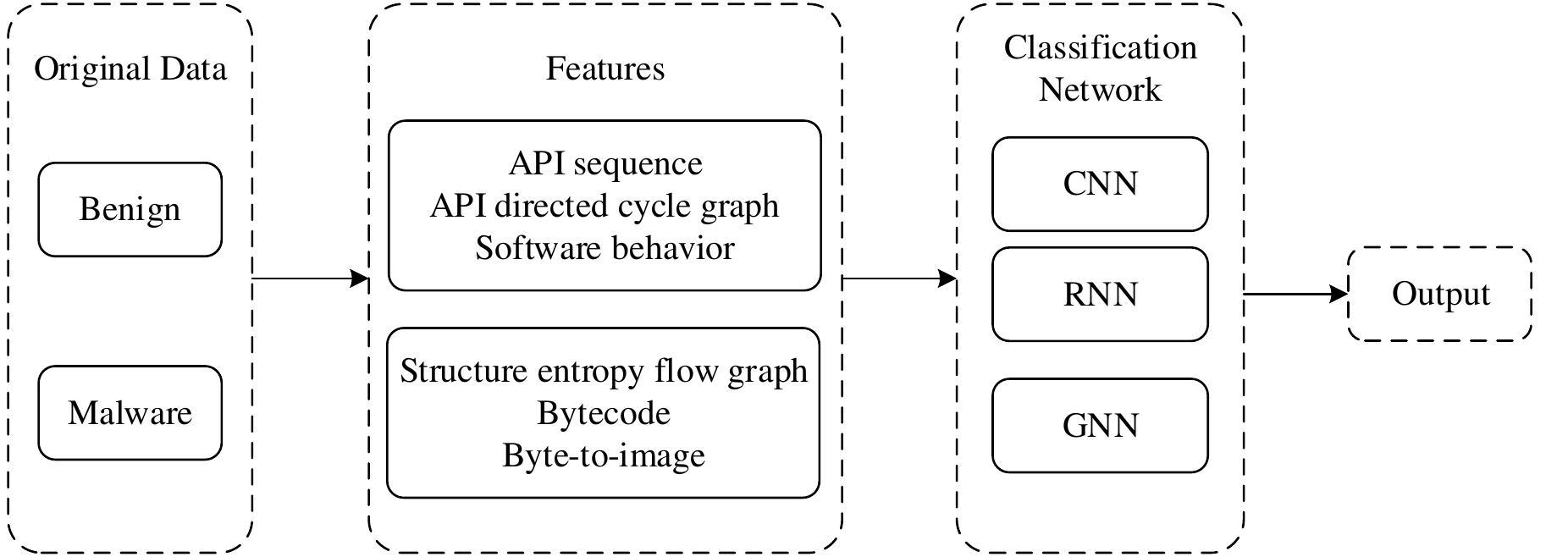}
    \caption{The main method of malware detection based on deep learning}
    \label{fig2}
\end{figure*}

Deep learning technology has been widely used in the field of malware detection. According to the type of selected features, the current main detection methods can be divided into dynamic feature-based and static feature-based, as shown in Figure\ref{fig2}.

Dynamic features are features obtained by using dedicated tools when PE files are run in an isolated environment such as a sandbox. There are mainly API sequences, API directed cycle graphs, and software behavior topology graphs. Among them, the detection method based on API sequence\cite{rosenberg2018generic} is to process the API calls of the software runtime into time series, and use the RNN network to analyze the context relationship of the sequence to detect malware. API-based directed cycle graph detection\cite{li2022intelligent} processes API calls into graph data and uses graph neural networks for classification. Topological graph detection based on software behavior\cite{liu2021mg} is to use the behaviors of file reading, memory reading, and file writing during software runtime as graph nodes, and build a topology graph according to the program running process, and then use graph neural network to classify.

Static features are features directly obtained from PE files, mainly including byte-to-image, structure entropy flow graph, bytecode, etc. Based on the byte-to-image detection method\cite{cui2018detection}\cite{tekerek2022novel}, the binary bytes of PE files are converted into images, and then classified by convolutional neural networks. The flow graph detection method based on structure entropy\cite{gibert2018classification} calculates the entropy value of each part according to the PE file structure, and uses wavelet transform to process the entropy value sequence into a manifold graph, and then uses the convolutional neural network to classify. Bytecode-based detection method (MalConv)\cite{raff2017malware} firstly performs word embedding on binary bytes of PE files, and then uses the convolutional neural network for classification. Compared with the detection method based on byte-to-image, this detection method both uses the binary bytes of the PE file as the original data, but there are two differences. First, MalConv performs word embedding on binary bytes, rather than converting them to images, making it necessary to process sequences of file size lengths\cite{raff2021classifying}. Specifically, when the input PE file is 2 MB, this detection method needs to process a sequence of about 2,000,000 steps. Therefore, this detection method consumes more resources than the detection method based on byte-to-image. In addition, the convolutional neural network layer used by MalConv is shallow, and the input volume of MalConv is too large. In order to reduce resource occupation, a deeper network layer cannot be used.

In practical applications, to avoid malware evading the detection model, it is necessary to combine a variety of static feature detection with dynamic feature detection. However, dynamic feature detection has problems such as long detection time, malware deliberately evading sandboxes, and high resource consumption. Therefore, when faced with a lot of malware in the external environment, static detection is the first and most important line of defense.

\subsection{Related Work}
For the adversarial malware attack of the malware byte-to-image detection model, Liu et al. \cite{liu2019atmpa} first proposed ATMPA, and directly used the FGSM(Fast Gradient Sign Method)\cite{goodfellow2014explaining}, DeepFool\cite{moosavi2016deepfool}, and C\&W\cite{carlini2017towards} methods to generate adversarial malware. It has been verified that the generated adversarial malware can obtain a very high rate of model misclassification, but the malware has a semantic feature, which directly adds perturbations to bytes of the malware, resulting in the generated adversarial malware losing its malicious function and executability. Therefore, Khormali et al. \cite{khormali2019copycat} propose an improvement to generate non-executable adversarial malware using ATMPA, take the non-executable adversarial malware as the perturbation, and append the perturbation to the end of the malware to generate the final adversarial malware, making the final adversarial malware adversarial, executable, and malicious. However, when this method generates adversarial malware, it needs to add a perturbation of the size of the malware, which will cause the generated adversarial malware to lose its concealment ability.

Benkraouda et al. \cite{benkraouda2021attacks} injected perturbation by inserting empty assembly instructions (such as NOP instructions) and combined the C\&W method to generate adversarial malware. The adversarial malware generated by this method has the slightest difference from the malware, but there are problems such as excessive perturbation injection restrictions and the high cost of generating adversarial malware.

Xiao Mao et al. \cite{mao2022adversarial} proposed a legal injection method (Bytecode Attack Remained Availability and Functionality, BARAF) that preserves the executable and functionality of the adversarial malware, adds specific bytes (such as 0x00, 0xFF) to the malware to destroy the texture feature of the sampled gray image, and finally successfully generates the adversarial malware. This method can generate executable and malicious adversarial malware with less perturbation. However, this study did not pay attention to the structure, gradient, output and other information of the attack target model, resulting in a low success rate of the generated adversarial malware to deceive the target model.

\section{Method}
This section first introduces the rule settings for adversarial malware adversarial attacks, then describes the adversarial malware generation method FGAM proposed in this paper, and finally introduces the functionality-preserving manipulations, which ensures that adversarial malware is malicious and executable.

\subsection{Malware Adversarial Attacks}
\subsubsection{Attack settings}
First, the ultimate goal of the attacker in this paper is to generate adversarial malware so that it can evade the malware detection model, i.e. the adversarial malware is classified as benign by the detection model. Second, the attacker must ensure that the adversarial malware has executable, malicious, and concealment capabilities. Finally, in this paper, the ability of the attacker is limited as follows. The attacker can only input the malware detection model, and obtain the gradient information and output results of the detection model. It is forbidden for attackers to pollute the data set of the model, obtain the data set of the model, change the network structure of the model, and inject network backdoors.

\subsubsection{Attack target}
\begin{figure*}[hbtp]
    \centering
    \includegraphics[scale=0.65]{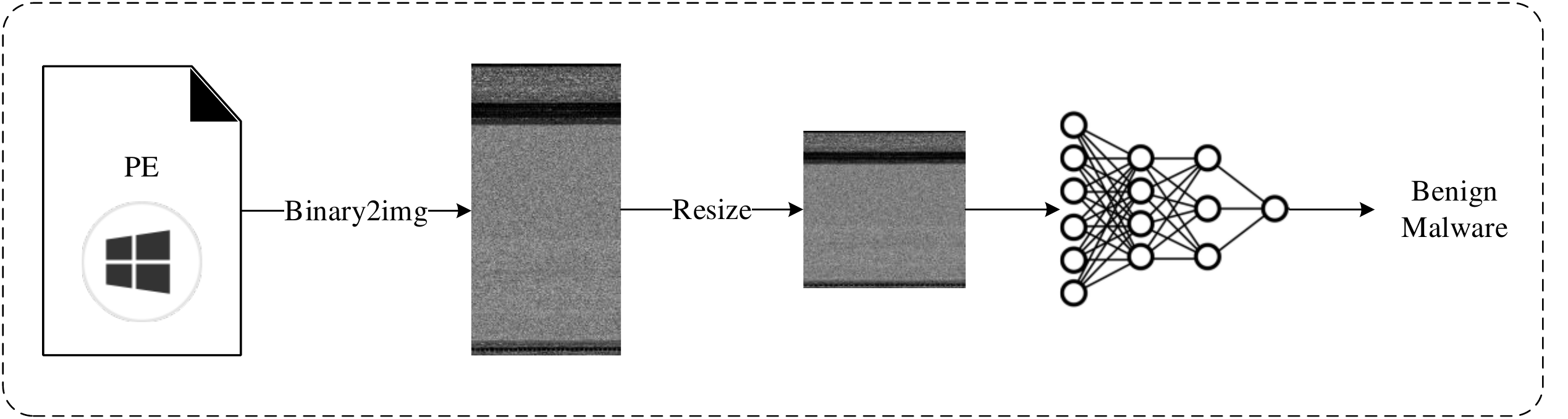}
    \caption{The process of malware detection method based on byte-to-image}
    \label{fig3}
\end{figure*}

The target of the adversarial attack in this paper is the malware detection model based on byte-to-image, and its complete detection process is shown in Figure\ref{fig3}. First, the Windows PE file is converted into a grayscale image through the Binary2img algorithm. Due to the different sizes of the PE files, the size of the converted grayscale image is also different. In order to facilitate the training of the neural network, the image needs to be adjusted to a uniform size. The algorithm used for Resize is the bilinear interpolation method. The adjusted image is input into the neural network, and the classification result of the detection model for the PE file can be obtained.

The specific content of the Binary2img algorithm is shown in Algorithm 1. When inputting, in order to preserve the characteristics of the PE file to the greatest extent, the width of the grayscale image needs to be set according to the size of the PE file. The corresponding relationship is shown in Table\ref{tab1}. This correspondence comes from the Malimg[21] dataset, but the original correspondence does not consider PE file over 1 MB. In this paper, the correspondence is extended to 15 MB according to the requirements. After inputting the binary file, first get its file size, and obtain the corresponding grayscale image width and height (line1, line2, line3) according to the file size. We convert the 8-bit unsigned binary number in the binary file to a decimal number between 0 and 255, and get the data list of the binary file, and get the length of the list (line4, line5). We calculate the size difference between the list and the image array, and perform zero-padding operations (line6, line7, line8, line9). Resize the list to an m*n image array, the target grayscale image (line11).
\begin{table}[!htb]\centering
    \caption {Misclassification rate and cost time}
    \label{tab1}
    \resizebox{.85\columnwidth}{!}{
        \begin{tabular}{llll}
            \toprule  %
            File size & Width & File size   & Width \\
            \midrule  %
            0-10 KB   & 32    & 100-200 KB  & 384   \\
            10-30 KB  & 64    & 200-500 KB  & 512   \\
            30-60 KB  & 128   & 500-1024 KB & 768   \\
            60-100 KB & 256   & 1-15 MB     & 1024  \\
            \bottomrule %
        \end{tabular}
    }
\end{table}

\begin{algorithm}[hbt]
    \caption{Binary2img}
    \begin{algorithmic}[1]
        \REQUIRE ~~ \\
        $x$, the input PE binary sample; \\
        $table$, the correspondence table between file size and grayscale image width;
        \ENSURE  ~~ \\
        $img$,the byte-based grayscale image;\\
        $m$, the width of the grayscale image;\\
        $n$, the height of grayscale image;\\
        \STATE $size=getsize(x)$
        \STATE $m \leftarrow table(size)$
        \STATE $n \leftarrow [size/m]$
        \STATE $list_{Dec} \leftarrow x$
        \STATE $length \leftarrow len(list_{Dec})$
        \STATE $temp \leftarrow m*n-length,i \leftarrow 0$
        \WHILE {$i<temp$}
        \STATE $list_{Dec}[length +i] \leftarrow 0$
        \STATE $i \leftarrow i+1$
        \ENDWHILE
        \STATE $img[m][n] \leftarrow reshape(list_{Dec})$
        \RETURN $img,m,n$
    \end{algorithmic}
\end{algorithm}

\subsection{FGAM}
This section first describes the problem of generating adversarial examples by mathematical methods as a minimization problem, and gives the objective function of the minimization problem, and then introduces the solution method of this objective function and the functionality-preserving manipulations used in FGAM.

\subsubsection{Problem function}
Let the malware detection model be $f$, the output label of the model be $y$, the input is $x$, and the output of the model $f(x)$ (the malicious score) range from [0, 1]. In this paper, the classification threshold of malware is set to 0.5, and the relationship between the model input and output is shown in Equation 1. According to the definition of adversarial samples, let the adversarial malware be $x'$ and the perturbation is $s$, then inject the perturbation into the original sample $x$ to get the adversarial malware $x'$ (Equation 2). If the generated adversarial malware can deceive the detection model, the requirements of Equation 3 need to be satisfied. Therefore, the problem of generative adversarial examples in this paper can be described as fixing the magnitude of the injected perturbation and updating the perturbation through gradient signs to minimize the maliciousness score of adversarial malware (Equation 4). The perturbation injection amount is jointly determined by the file size and the perturbation injection ratio (Equation 5). The perturbation injection amount of different files are different, so the injection ratio is used to represent the perturbation injection amount.
\begin{equation}
    y\left\{\begin{array}{l}\text { benign, } f(x)<0.5 \\\text { malware, } f(x) \geq 0.5\end{array}\right.
\end{equation}
\begin{equation}
    x'=H(s+x)
\end{equation}
\begin{equation}
    f(x')=f(H(s+x))<0.5
\end{equation}
\begin{equation}
    {\rm minimize} \\\ f(x')=f(H(x+s)) \\\ update \\\ s \\\ with \\\ gradient
\end{equation}
\begin{equation}
    s_{size}=rate*x_{size}
\end{equation}
\subsubsection{FGAM}
\begin{figure*}[hbtp]
    \centering
    \includegraphics[scale=0.7]{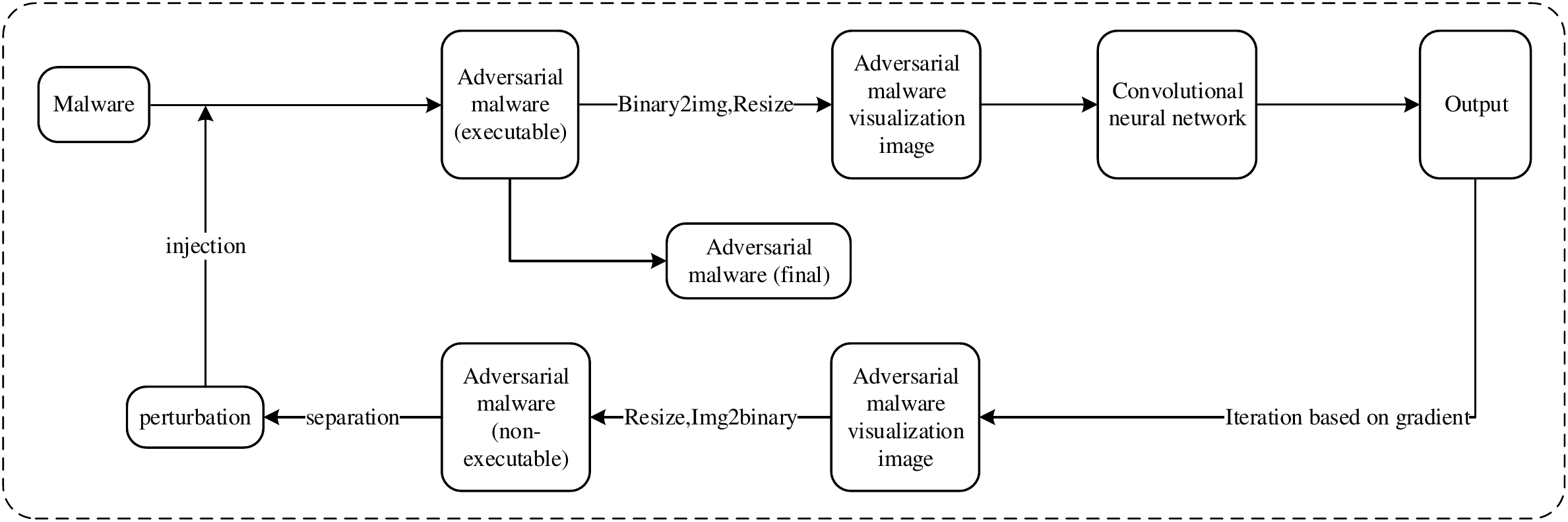}
    \caption{The adversarial malware generation process}
    \label{fig4}
\end{figure*}

The process of FGAM generating adversarial malware is shown in Figure\ref{fig4}. First, inject perturbation into the malware to generate adversarial malware, input it into the detection model, and generate a visual image of the adversarial malware according to the output results and gradient information. Convert the image into adversarial malware. The adversarial malware is not executable at this time, so it is necessary to separate the perturbation from it, and inject the separated perturbation into the malware again to generate executable adversarial malware. The above process is repeated, and if the generated executable adversarial malware can fool the detection model. That is, the output result of the model is benign software, the process will be stopped.

The specific algorithm of FGAM is shown in Algorithm 2. First, the injected perturbation amount is determined according to the original malware size and injection rate, and the injected perturbation is initialized to random values (line1-line3). The perturbation is injected into the malware to generate adversarial malware, and the adversarial malware is adjusted into a visualized image. The iterative descending speed and model output initial value is given, so that it can satisfy the iteration start conditions (line5-line8). The iterative loop begins, first to judge whether the adversarial malware successfully deceives the detection model, and the classification threshold of the detection model in this paper is detailed in Section 3.2.1. If the deception is successful, stop the iteration and return the adversarial malware and the number of iterations (line9-ine10). If it fails, the adversarial visualization images are generated repeatedly and iteratively using the fast gradient sign method to ensure that it can fool the detection model (line11-line15). Converting adversarial visualization images into adversarial malware, although adversarial malware can deceive the detection model, but lose its executable and malicious. Therefore, it is necessary to separate adversarial perturbations from them and inject the separated perturbations into the original malware to generate executable and malicious adversarial malware. But its adversarial weakens, and it needs to be tested to whether it can fool the detection model. (line17-line18). At the end of each iteration, the malicious score of the adversarial malware (executable) is recorded, and the malicious score curve is fitted by the least square method, and the slope of the curve is used to draw the score decline curve, and the slope of the score decline curve is used as the decline speed of the malicious score. (line19). The decline speed of malicious scores can be used to judge whether the iteration process is in oscillation. Falling into oscillation does not mean that effective adversarial malware cannot be generated by continuous iteration, but the purpose of the FGAM proposed in this paper is to generate adversarial malware quickly, so it is necessary to reduce the number of iterations used and avoid the generation process falling into oscillation.
\begin{algorithm}[hbt]
    \caption{FGAM}
    \begin{algorithmic}[1]
        \REQUIRE ~~ \\
        $x$,  the input binary malware sample; \\
        $y$,the targets associated with x;\\
        $\theta $, the parameters of the model;\\
        $\varepsilon  $, the perturbation step;\\
        $rate$,the proportion of injected perturbation;\\
        $model$,the malware detection model;\\
        $T$, the maximum number of iterations;
        \ENSURE~~ \\
        $x'$,the adversarial malware;\\
        $t$, the actual number of iterations;
        \STATE $size \leftarrow getsize() $
        \WHILE {$i<size*rate$}
        \STATE $s_i \leftarrow random(0,255),i \leftarrow i+1$
        \ENDWHILE
        \STATE $speed \leftarrow 1,model_{out} \leftarrow 1,t \leftarrow 0$
        \WHILE {$t<T$ \AND $speed>0.001$ \AND $model_{out}>0.5$}
        \STATE $x' \leftarrow inject \quad s \quad to \quad x$
        \STATE $x_{img} \leftarrow Rsize(Binary2img(x'))$
        \STATE $model_{out} \leftarrow model(x_{img}),t \leftarrow t+1$
        \IF {$model_{out}>0.5$}
        \STATE $model_{advout} \leftarrow 1$
        \WHILE { $model_{advout}>0.5$}
        \STATE$x_{advimg} \leftarrow x_{img}  +\varepsilon sign(\nabla_{x_{img}}J(\theta,x_{img},y) )$
        \STATE$model_{advout} \leftarrow model(x_{advimg})$
        \STATE$x_{img} \leftarrow x_{advimg} $
        \ENDWHILE
        \STATE $x' \leftarrow Img2binary(Resize(x_{advimg}))$
        \STATE$ s \leftarrow separate \quad s \quad from \quad x'$
        \STATE$ speed \leftarrow LeastSquares(model_{out})$
        \ENDIF
        \ENDWHILE
        \RETURN $x',t$
    \end{algorithmic}
\end{algorithm}

\subsubsection{Functionality-preserving manipulations}
\begin{figure}[hbtp]
    \centering
    \includegraphics[scale=0.6]{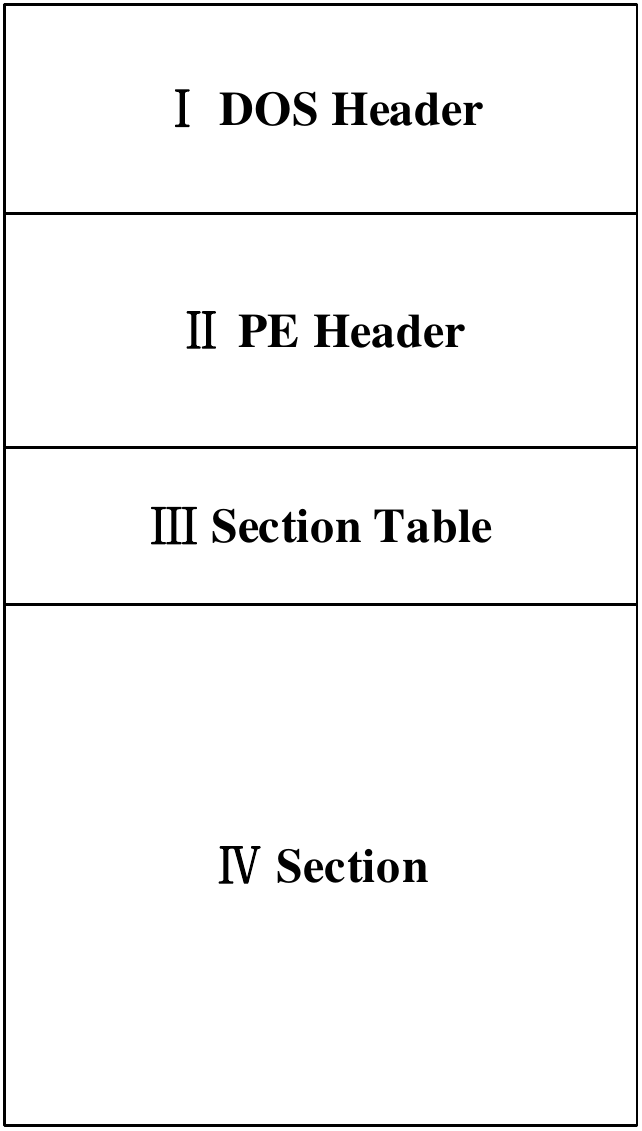}
    \caption{PE file structure simplified diagram}
    \label{fig5}
\end{figure}

Functionality-preserving manipulations are the key to ensuring that adversarial malware is executable and malicious, and its main principle is to inject perturbations into areas not used by program execution. A simplified diagram of the PE file structure is shown in Figure\ref{fig5}. We will introduce some function-preserving manipulations based on the PE structure. According to the injection location classification, the existing functionality-preserving manipulations are as follows. (1) Full the DOS header\cite{demetrio2021adversarial}. Since the MZ flag and the PE offset part of the DOS header cannot be changed, the perturbation can be injected into other parts of the DOS header to fill the space of the DOS header. (2) Extend DOS header\cite{demetrio2021adversarial}. By rewriting PE offset, inject a segment of byte perturbation between I and II, and ensure that it is aligned with the PE file, which is equivalent to increasing the space of the DOS header. (3) Injecting section\cite{demetrio2021adversarial}, by rewriting the offset address of the section, injecting a section perturbation before the first section of IV. Section perturbations can also be injected into other locations of the IV, but in order to avoid too much intrusion into the PE file, it is preferred to inject before the first section or after the last section. (4) Padding\cite{kolosnjaji2018adversarial}\cite{li2023gambd} Adding byte perturbation directly at the end of the PE file (after IV), this method is the least intrusive and the operation is the easiest.

In conclusion, functionality-preserving manipulations is about injecting perturbations into the PE file, and the injected perturbations will not be executed and will not affect the execution. However, different functionality-preserving manipulations have different injection locations and different intrusiveness. Most of the existing PE files cannot run in the DOS environment, and the content of the DOS header of the PE files is relatively fixed. When adversarial malware is generated using the full DOS header and extend DOS to inject perturbation, the DOS header part of the generated adversarial malware is quite different from that of benign software. Therefore, the adversarial malware generated by the above two methods have poor concealment and are not used in this paper. This paper takes injecting section and Padding at the end as injection operation.

\section{Experiment}
The experiment uses PyTorch as the deep learning framework, Python as the programming language, and uses the NVIDIA T4 (16G) graphics card to provide acceleration. The specific physical device is a high-performance server equipped with Centos7, the CPU is Intel(R) Xeon(R) Gold 5218, and it is equipped with 256G RAM.

\subsection{ Dataset and Models}
Since there is currently no open-source malware binary dataset, it needs to be collected through VirusShare\footnote{https://virusshare.com/}. In addition, there is currently no open-source malware detection model, so the malware detection model needs to be trained using the dataset in this paper, and the dataset is divided into the training set and the test set with a ratio of 7:3. The composition of this dataset is shown in Table\ref{tab2}.
\begin{table}[!htb]\centering
    \caption {Dataset Composition}
    \label{tab2}
    \resizebox{.95\columnwidth}{!}{
        \begin{tabular}{llll}
            \toprule  %
            Data    & Training set & Testing set & Source     \\
            \midrule  %
            Benign  & 9489         & 4068        & Windows    \\
            Malware & 8473         & 3632        & VirusShare \\
            \bottomrule %
        \end{tabular}
    }
\end{table}

The existing byte-to-image malware detection models have the same preprocessing process, but the difference lies in the convolutional neural network used for classification. After literature research, this paper selects the latest detection model as the attack target. Tekerek et al. \cite{tekerek2022novel} proposed to use DenseNet121 as the classification network. When training the network, a dynamic learning rate was used, with an initial learning rate of 0.1, and after every five iterations, the learning rate was reduced to 0.5 times the original. After 50 rounds, the best performance in the test set is selected as the final detection model of this paper. The model test results are shown in Table\ref{tab3}.

In addition, in order to evaluate the transferability of adversarial malware, it is necessary to train the transfer target model with the dataset in this paper. In this paper, MalConv\cite{raff2017malware} is selected as the transfer target model. The maximum input of the model is 15MB. After 50 rounds of training, the best test set is selected as the final model in this paper. The model test results are shown in Table\ref{tab3}.
\begin{table}[!htb]\centering
    \caption {Malware detection model testing results}
    \label{tab3}
    \resizebox{.95\columnwidth}{!}{
        \begin{tabular}{lll}
            \toprule  %
            Model                             & Accuracy & Auc     \\
            \midrule  %
            Attack target model (DenseNet121) & 94.12\%  & 97.69\% \\
            Transfer target model (MalConv)   & 94.18\%  & 98.12\% \\
            \bottomrule %
        \end{tabular}
    }
\end{table}

\subsection{Adversarial malware attack evaluation}
In Table\ref{tab2333}, we evaluate adversarial malware generated by existing work. We evaluate by the functionality of adversarial malware and the minimal proportion of injection perturbations. Functionality indicates whether the generated adversarial malware retains the functionality of the original malware. The minimum injection ratio represents the minimum perturbation injection required when adversarial malware is effective. Since the sizes of different files are different, the ratio of the perturbation amount to the original file is used to indicate the injection size. The adversarial malware generated by ATMPA loses functionality, and the amount of perturbation required for adversarial malware generated by COPYCAT is too large, so we use BARAF as a comparison work in this paper.
\begin{table}[!htb]\centering
    \caption {Assessment of existing methods}
    \label{tab2333}
    \resizebox{.80\columnwidth}{!}{
        \begin{tabular}{lll}
            \toprule  %
            Method                            & functionality & perturbation rate \\
            \midrule  %
            ATMPA\cite{liu2019atmpa}          & NO            & 100\%             \\
            COPYCAT\cite{khormali2019copycat} & YES           & 100\%             \\
            BARAF\cite{mao2022adversarial}    & YES           & 5\%               \\
            \bottomrule %
        \end{tabular}
    }
\end{table}

To verify the effectiveness of FGAM, the adversarial malware generated by this method are evaluated for adversarial attack capability. The experiment selects 500 malwares from the dataset as the original samples, and uses three methods such as FGAM, Random (injecting random perturbation), and BARAF \cite{mao2022adversarial} to generate adversarial malware. This paper evaluates the attack ability of adversarial malware by the misclassification rate (MR) of the model. The selected malware samples can be correctly classified by the detection model, so the initial misclassification rate of the model is 0. Therefore, the higher the misclassification rate of the model, the higher the success rate of adversarial attacks and the more effective the method of generating adversarial malware. The experimental results are shown in Table\ref{tab4}, where MR(0) represents the misclassification rate of the detection model when the number of iterations is 0. MR(20) indicates the misclassification rate of the detection model when iteratively 20 times, but the Random and BARAF methods cannot iterate, so there is no experimental result.
\begin{table}[!htb]\centering
    \caption {The model misclassification rate}
    \label{tab4}
    \resizebox{.95\columnwidth}{!}{
        \begin{tabular}{llll}
            \toprule  %
            Method & Inject operation  & MR(0) & MR(20)          \\
            \midrule  %
            \multirow{2}{*}{Random}
                   & Injecting section & 2.6\% & No iteration    \\

                   & Padding           & 2.8\% & No iteration    \\
            \multirow{2}{*}{BARAF\cite{mao2022adversarial}}
                   & Injecting section & 8\%   & No iteration    \\

                   & Padding           & 2.2\% & No iteration    \\
            \multirow{2}{*}{FGAM}
                   & Injecting section & 2.6\% & \textbf{87.2\%} \\

                   & Padding           & 2.8\% & \textbf{91.6\%} \\
            \bottomrule %
        \end{tabular}
    }
\end{table}

\begin{figure*}[hbtp]
    \centering
    \includegraphics[scale=0.75]{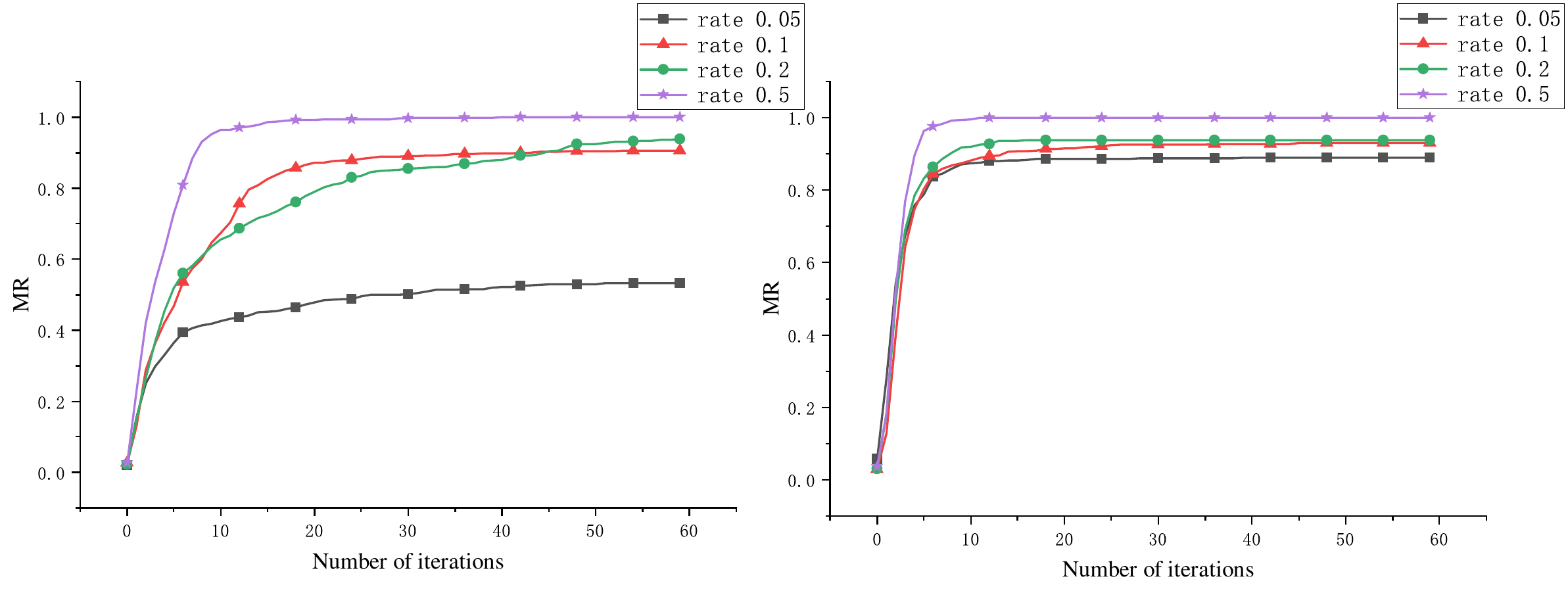}
    \caption{Model misclassification rate. The left picture is the experimental result of inject section, and the right picture is the experimental result of padding}
    \label{fig6}
\end{figure*}

The above experimental results show that FGAM enhances perturbation adversarial ability through iteration, and improves the success rate of generating adversarial malware, but the adversarial malware generated under different injection rates and injection methods are different. To further study this problem, we set up various control experimental groups, selected 500 malware samples, and used various functionality-preserving manipulations and injection rates to generate adversarial malware. The experimental results are shown in Figure\ref{fig6}. The injection rates were set to 5\%, 10\%, 20\%, and 50\%, respectively. The experimental results show that at the beginning of the iterative process, the number of iterations is the decisive factor to generate effective adversarial malware, and the effect of increasing the injection rate is not significant at this time. However, as the number of iterations increases, its marginal effect diminishes, and the injection rate becomes the decisive factor. In short, the injection rate determines the upper bound on the quality of adversarial examples. When increasing the number of iterations fails to generate adversarial malware effectively, it is necessary to consider increasing the injection rate. But there are also limitations on the injection rate, which are described in Section 4.3.

\subsection{Static information statistics}
\begin{figure}[hbtp]
    \centering
    \includegraphics[scale=0.3]{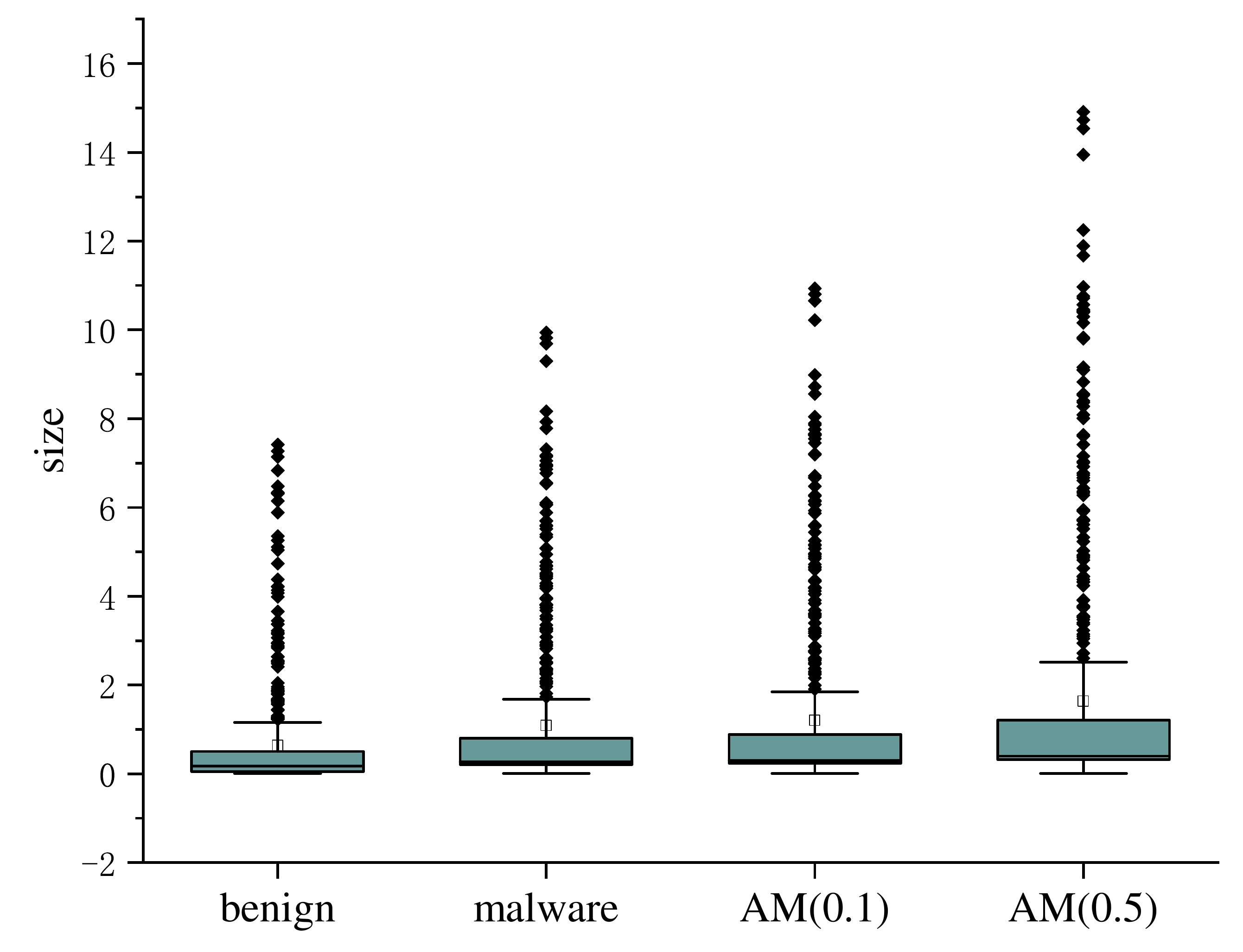}
    \caption{Statistics of software file size}
    \label{fig7}
\end{figure}

To verify the concealment ability of adversarial examples generated by FGAM under different injection rates. The file size and entropy values of benign software, malware, adversarial malware (0.1), and adversarial malware (0.5) were counted respectively. The statistical sample size is 500. The statistical results of file size are shown in Figure\ref{fig7}. The statistical results show that there is no significant difference in the size distribution of files with the increase of the injection rate. In conclusion, the file size will not be the limit of the injection rate.

\begin{figure}[hbtp]
    \centering
    \includegraphics[scale=0.3]{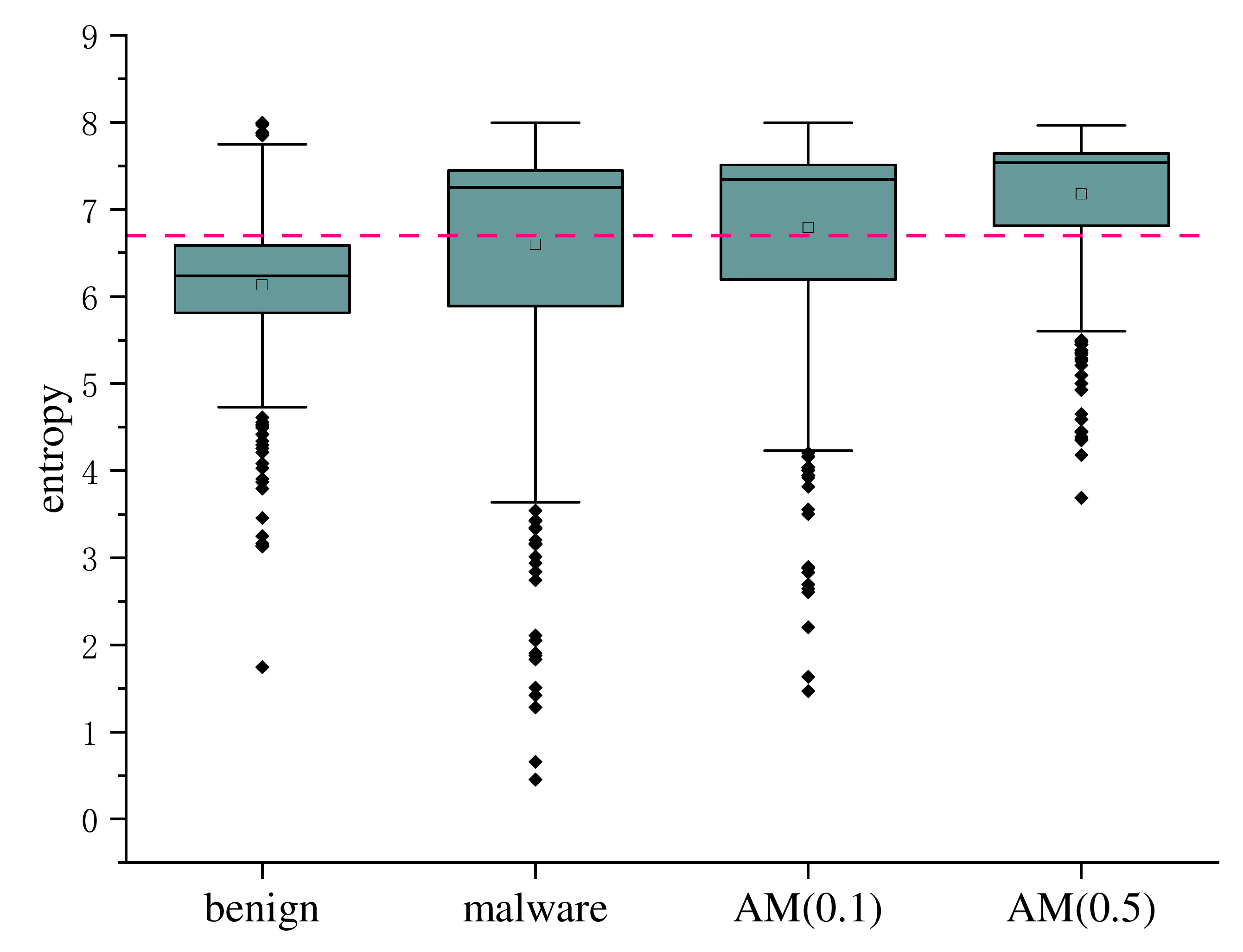}
    \caption{Statistics of software entropy}
    \label{fig8}
\end{figure}

The statistical results of software entropy are shown in Figure\ref{fig8}. The results show that the entropy distributions of malware and benign software have overlapping areas. When the injection rate is 10\%, the entropy distribution of adversarial malware shifts upward, but there is still an overlap region with the distribution of benign software. However, when the injection rate is 50\%, the entropy distribution of adversarial malware shifts further up, and there is a clear demarcation from the distribution of benign software. However, when the injection ratio is 50\%, the entropy distribution of adversarial malware is further shifted upward, and there is a clear distinction with the distribution of benign software.

Therefore, there is a limit to the perturbation injection rate, and injecting too much perturbation will lead to a significant increase in file entropy. When generating adversarial examples, increasing the injection rate can enhance the adversarial ability of the adversarial examples, making it easier to fool the detection model. However, at the same time, the file entropy increases significantly, and there is a clear distinction between adversarial malware and benign software in entropy, which leads to the failure of the adversarial malware deception detection model.

\subsection{Adversarial attack transfer capability}
To evaluate the transfer ability of adversarial malware, the adversarial malware generated by FGAM are input into the transfer target model, and the misclassification rate (MR) of the transfer target model is used to evaluate the transfer ability of the adversarial malware. In this paper, MalConv is selected as the transfer attack target model. Both MalConv and the attack target model in this paper use bytes as the original features. The training details and test results of the target transfer model are shown in 5.1. First, evaluate the transferability of adversarial examples generated by existing methods. In the experiment, the functionality-preserving manipulation was selected as the inject section, and the injection rate was 10\%. The experimental results are shown in Table\ref{tab5}. The experimental results show that the adversarial malware generated by FGAM have a strong adversarial attack ability to the malware detection model based on byte-to-image. However, its attack ability on the transfer model is similar to the adversarial malware generated by Random.
\begin{table}[!htb]\centering
    \caption {The model misclassification rate}
    \label{tab5}
    \resizebox{.95\columnwidth}{!}{
        \begin{tabular}{lcc}
            \toprule  %
            Sample     & \makecell{ Attack target model          \\ (DenseNet121)} & \makecell{Transfer target model \\ (MalConv) } \\
            \midrule  %
            AM(Random) & 2.6\%                          & 25.6\% \\
            AM(BARAF)  & 8\%                            & 29\%   \\
            AM(FGAM)   & 90.6\%                         & 25.8\% \\
            \bottomrule %
        \end{tabular}
    }
\end{table}

In order to explore whether the transfer ability of adversarial malware is related to different functionality-preserving manipulations, different injection rates, and different adversarial strengths. In the experiment, 500 malware were selected, and under different conditions, FGAM was used to generate adversarial malware, and the transfer ability of the adversarial malware was tested. When the inject section is used as the functionality-preserving manipulation, the experimental results are shown in Table\ref{tab6}. The different adversarial strengths represent different model thresholds for stopping iteration in the FGAM algorithm. The original threshold is 0.5, and the enhancement threshold is set to 0.9. The experimental results show that enhancing the malicious score and injection rate cannot improve the transfer attack ability of adversarial malware. Conversely, reducing the injection rate can improve the ability of adversarial malware transfer attacks.
\begin{table}[!htb]\centering
    \caption {The model misclassification rate. When the functionality-preserving manipulation is injecting section}
    \label{tab6}
    \resizebox{.95\columnwidth}{!}{
        \begin{tabular}{lcc}
            \toprule  %
            Sample       & \makecell{ Attack target model          \\ (DenseNet121)} & \makecell{Transfer target model \\ (MalConv) } \\
            \midrule  %
            Malware      & 0                              & 0      \\
            AM(10\%,0.5) & 90.6\%                         & 25.8\% \\
            AM(5\%,0.5)  & 53.2\%                         & 28.6\% \\
            AM(20\%,0.5) & 93.8\%                         & 24.6\% \\
            AM(50\%,0.5) & 100\%                          & 22\%   \\
            AM(10\%,0.9) & 73.4\%                         & 25.6\% \\
            \bottomrule %
        \end{tabular}
    }
\end{table}
When padding is used as the functionality-preserving manipulation, the experimental results are shown in Table\ref{tab7}. The experimental results show that the transfer ability of adversarial malware is completely lost, which is consistent with the interpretability conclusion of MalConv. Demetrio et al. \cite{demetrio2019explaining} conducted an interpretability analysis of MalConv by CAM(Class Activation Mapping)\cite{zhou2016learning}, and the results showed that the head position information of the PE file has a significant impact on the classification results of the MalConv model. Therefore, when padding is used as the functionality-preserving manipulation, the perturbation is fully injected into the end of the malware, and the header information of the malware does not change, so the ability of the adversarial malware to deceive the model is weakened.

The perturbation injection position of the inject section is close to the head of the PE file, so the generated adversarial malware can transfer attacks. In PE files, sections occupy most of the file content. So when the injection section position is before the first section, the injection position is close to the head of the PE file. When the injection section position is before the middle section, the injection position is the middle position of the PE file.
\begin{table}[!htb]\centering
    \caption {The model misclassification rate. When the functionality-preserving manipulation is padding}
    \label{tab7}
    \resizebox{.95\columnwidth}{!}{
        \begin{tabular}{lcc}
            \toprule  %
            Sample       & \makecell{ Attack target model         \\ (DenseNet121)} & \makecell{Transfer target model \\ (MalConv) } \\
            \midrule  %
            Malware      & 0                              & 0     \\
            AM(10\%,0.5) & 93\%                           & 0.6\% \\
            AM(5\%,0.5)  & 89\%                           & 0.6\% \\
            AM(20\%,0.5) & 93.8\%                         & 0.6\% \\
            AM(50\%,0.5) & 100\%                          & 0.8\% \\
            AM(10\%,0.9) & 87\%                           & 1\%   \\
            \bottomrule %
        \end{tabular}
    }
\end{table}

The experimental results of the adversarial malware transfer attack generated by the above two functionality-preserving manipulations show that when the perturbation injection position is close to the PE file header, the adversarial malware has the ability to transfer attack; When the perturbation is injected at the end of the PE file, the transfer ability of the adversarial malware is almost lost, and the experimental results are mutually confirmed with the interpretability conclusion of MalConv. The experimental results in Section 4.2 show that under the same perturbation injection rate and number of iterations, the padding operation is more efficient to generate adversarial malware. Therefore, the impact of PE file header information on the byte-to-image detection model is not as significant as MalConv.

In conclusion, both models are end-to-end malware detection models, but the decision-making basis of the two is not the same. First, the adversarial malware generated by FGAM have a deception rate of about 90\% for the byte-to-image detection model, but the deception effect on the MalConv model is similar to the adversarial malware generated by Random, and there is no obvious advantage. Second, when the two models detect the same PE file, the classification decisions rely on different regions. Therefore, the two detection models can be integrated through ensemble learning, etc., to improve the accuracy and robustness of the malware end-to-end detection model.

\section{Conclusion}
In this paper, we propose FGAM, a method for fast generation of adversarial malware. This method uses iterative perturbation to enhance perturbation adversarial ability, and the enhancement of perturbation adversarial ability can reduce perturbation injection amount and improve the success rate of adversarial attacks. The goal of the iteration is to minimize the adversarial malware score. FGAM detects the rate of decline of malicious scores, stops meaningless iterations in advance, and limits the number of iterations to a certain range. In addition, perturbation injection through function-preserving manipulations ensures that adversarial malware is malicious and executable. In the experiment, the static information of the adversarial malware is counted, and it is proposed that the perturbation injection rate has an entropy limit, and the perturbation injection rate and the concealment ability of the adversarial malware are mutually restricted. Finally, under different functionality-preserving manipulations and different perturbation injection rates, the efficiency of the method to generate adversarial examples and the transfer attack ability of adversarial examples are analyzed.

The work in this paper mainly realizes the rapid generation of adversarial malware, but there are still some shortcomings, which will be improved in future work. First, increase the variety of functionality-preserving manipulations. The generated adversarial malware consists of malware with adversarial perturbations, where the input of the adversarial perturbation causes the detection model to produce false outputs. Therefore, increasing the diversity of functionality-preserving manipulations can make it more challenging to separate adversarial perturbation from malware, and enhance the ability of adversarial malware attacks. Second, combine the sensitivity analysis with perturbation injection. The experimental results show that different injection locations have different efficiencies in generating adversarial examples. Therefore, sensitivity analysis can be used to obtain the region in the sample that has a significant influence on the classification decision, and then inject perturbation into this region to improve the efficiency of generating adversarial malware.

\bibliographystyle{IEEEtran}
\bibliography{acmart}
\end{document}